\documentclass[epj]{svjour}
\usepackage{epsfig}

\begin{document}
\newcommand{\bq}{\begin{equation}}
\newcommand{\eq}{\end{equation}}
\newcommand{\bt}{\begin{tabular}}
\newcommand{\et}{\end{tabular}}
\newcommand{\ba}{\begin{array}}
\newcommand{\ea}{\end{array}}
\newcommand{\refe}[1]{(\ref{#1})}
\newcommand{\dst}{\displaystyle}
\newcommand{\xmed}[1]{\mbox{$\frac{#1}{2}$}}

\title{Thermal diffusion of sine-Gordon solitons}

\author{Niurka R.\ Quintero\inst{1} \and Angel S\'anchez\inst{1}
\and Franz G.\ Mertens\inst{2}}

\institute{Grupo Interdisciplinar de Sistemas
Complicados (GISC), Departamento de Matem\'aticas, 
Universidad Carlos III de Madrid,\\
Edificio Sabatini,
Avenida de la Universidad 30, E-28911 Legan\'{e}s, Madrid, Spain\\
\email{kinter@math.uc3m.es,\,anxo@math.uc3m.es}
\and Physikalisches Institut, Universit\"at Bayreuth, 
D-95440 Bayreuth, Germany\\
\email{franz.mertens@theo.phy.uni-bayreuth.de}}

\date{Received: \today / Revised version: }  

\abstract{ 
We analyze the diffusive motion of kink solitons governed by the thermal
sine-Gordon equation.  We analytically calculate the correlation
function of the position of the kink center as well as 
the diffusion coefficient, both up to second-order in temperature. 
We find that the kink behavior is very similar to that obtained in 
the overdamped limit: There is a quadratic dependence on temperature 
in the diffusion coefficient that comes from the interaction among the
kink and phonons, and the average value of the wave function increases
with $\sqrt{t}$ due to the variance of the centers of individual
realizations and not due to kink distortions. These analytical results 
are fully confirmed by numerical simulations. 
} 

\maketitle

\section{Introduction}
\label{intro}
As a key subject within nonlinear science, the dynamics of emergent, 
coherent structures (solitons, vortices, etc) has been a
research topic that has attracted very much attention in the past 
quarter century \cite{scott}. 
One question, extensively investigated in the literature 
\cite{bass,kiv,oldanx,kvbook,yuripr,anx} is 
the following: Is, and if so, how is the motion and the shape of those 
excitations modified by the presence of small perturbations? Indeed, 
when applied to physical situations of interest, nonlinear models must
incorporate additional terms, such as damping, constant or 
periodic external forces, or noise, to name a few.
Among those, stochastic perturbations are very much of interest in view
of their highly non trivial effects on nonlinear systems \cite{gsbook},
and a great deal of work has been devoted to them \cite{bass,oldanx,kvbook}.
In particular, of the very many nonlinear models applied to physical
problems, the sine-Gordon (sG) equation has been considered in much 
detail in this context, as it applies to, e.g., 
propagation of ultra-short optical pulses in resonant
laser media \cite{lamb}, a unitary theory of elementary particles
\cite{skyrme1,skyrme2,enz,raja}, propagation of magnetic flux in Josephson
junctions \cite{Barone}, transmission of ferromagnetic waves \cite{feld},
epitaxial growth of thin films \cite{CW,krug,us}, motion of dislocations in
crystals \cite{frenkel,Nabarro}, flux-line unlocking in type II
superconductors \cite{nuevo}, or DNA dynamics \cite{eng,DNA,yaku},
situations in which noise (of different origins) can play, and 
often does, a crucial role. As an example, let us mention the 
recent work on long Josephson junctions reported in \cite{exp},
where the authors calculated the escape rate from the zero-voltage state
induced by thermal fluctuations, obtaining very satisfactory results 
compared with the experimental ones. 
 
Specifically, 
this work is devoted to the study of the diffusive dynamics of 
sG kink solitons subjected to a thermal bath, as given by the 
stochastically perturbed, damped sG equation 
\begin{equation}
\phi_{tt} - \phi_{xx} + \sin(\phi) = -\alpha \phi_{t} + f(x,t,\phi_,...),
\label{ecua1}
\end{equation}
where $-\alpha \phi_{t}$ is a damping term with a dissipation 
coefficient $\alpha$, and $f(x,t,\phi_,...)$ is a thermal (gaussian)
noise term fulfilling
\begin{equation}
\begin{array}{l}
f(x,t,\phi_,...) = \sqrt{D} \,\ \eta(x,t), \quad 
\langle\eta(x,t)\rangle = 0, \\
\\
\langle\eta(x,t) \eta(x',t')\rangle = \delta(x-x') \delta(t-t'), 
\label{ecua2}
\end{array}
\end{equation}
where $\sqrt{D}$ is related to temperature through
the fluctu\-ation-dissipation theorem $D=2 \alpha k_{b} T$, 
$k_{b}$ being the Boltzmann constant 
and $T$ the temperature. 

To our knowledge, the first results on problems directly related to 
the one we deal with here were obtained by 
Joergensen {\em et al.} \cite{joer}, 
who performed experiments on Josephson junctions
and presented a derivation of the diffusion constant for kinks.
Subsequently, 
Kaup and Osman \cite{kaup} studied, in a more rigorous way,
the motion of damped 
sG kinks, driven by a constant force, in the presence of thermal 
fluctuations by using a singular perturbation expansion. They analyzed the 
temperature effect on the mean velocity of the kink and also 
the changes in the shape of the kink. In addition,
they calculated the diffusion coefficient of the kink up to first-order 
in temperature and the energy values corresponding to the 
translational ($E_{T}= k_{b} T/2$) and radiational ($E_{R}= k_{b} T$) modes. 
These values of the energy have been also obtained by Marchesoni 
\cite{march}, who applied the McLaughlin and Scott approach \cite{McL} to 
investigate kink motion under thermal fluctuations (see \cite{bass,kiv,kvbook}
for reviews). 

For the sake of completeness, let us mention work done along a different
line, namely that devoted
to the diffusive motion of the kink in equilibrium with phonons 
in {\em isolated sG systems}
(possibly perturbed) \cite{yuripr,Wsch,March2,March}.
In this case, the kink 
diffusive motion is characterized by two diffusion coefficients.
The first one of 
them is proportional to $T^{2}$ and is related to the anomalous diffusion,
that arises from the phase shifts of kinks colliding with phonons and 
takes place on a short time scale in which 
the collision among kink and wave packet is elastic; the kink 
retains the same velocity after the collision 
(non-dissipative diffusion) and suffers only a spatial shift. 
However, for large times and in slightly perturbed sG 
systems, this interaction is nonlinear and 
becomes inelastic and the velocity of the kink changes after the 
collisions \cite{Wada}. This diffusive regime is called viscous and 
has a diffusion coefficient proportional to $T^{-1}$. 
The diffusion of the kink when the low energy excitations are 
represented by breathers has also been studied,
and in \cite{Niko} it has been 
demonstrated that both descriptions (breathers or phonons) 
are equivalent and give rise to the same diffusion coefficient in the 
anomalous regime.

In any event, we want to stress that, 
although in this type of diffusion problem there are many open 
questions \cite{Wada}, we will concern ourselves with the other kind 
of diffusion problem, in which the phonons appear as a consequence 
of {\em an external heat bath}, represented by
white noise correlated in space and in time and the damping is 
included explicitly {\em \`a la} Langevin. The main aim of this
work is to extend a previous study of ours about the overdamped limit
of sine-Gordon kink diffusion \cite{nos} to the more physical and 
general case of the underdamped dynamics 
(i.e., with finite dissipation coefficient). 
As we will see below, the general perturbative approach 
\cite{raja} we resorted to in the overdamped case can also be applied,
albeit with more difficulties, to the underdamped problem. The 
corresponding theoretical analysis is presented in Sec.\ 2, where 
we obtain explicit expressions for the long-time diffusive dynamics
of kinks up to second-order in temperature, thus going beyond the 
currently available knowledge. The accuracy and importance of the 
new terms is assessed by numerical simulations in Sec.\ 3: we will 
see there that the quadratic corrections are in good agreement with 
the simulations and, most importantly, that they must be taken into
account even for not so large temperatures.
Finally, in the conclusions we summarize the main results of this 
work, comparing the underdamped and overdamped dynamics of the sG
equation and discussing other related questions.

\section{Analytical results}

We begin by briefly reviewing the basic results we need for our 
analytical approach. We will concern ourselves with the perturbation 
effect on the kink solutions of the unperturbed
sG equation, whose static form is
\begin{equation}
\phi_{0}(x,t)= 4 \,\ \mbox{arctan}[\exp(x)].
\label{ap1}
\end{equation}
Small perturbations over this equation can be treated by calculating the
spectrum of linear excitations around the kink solution \cite{bis}:
To this end, we write 
\begin{equation}
\phi(x,t)=\phi_{0}(x) + \psi(x,t), \,\ \psi(x,t) << \phi_{0}(x),
\label{ap2}
\end{equation}
substitute in (\ref{ecua1}) (with $\alpha=D=0$)
and linearize
around $\phi_{0}(x)$, arriving at the following equation for $\psi(x,t)$:
\begin{equation}
\psi_{tt} = \psi_{xx} - \Big[1-\frac{2}{\cosh^{2}(x)}\Big] \psi.
\label{ap3}
\end{equation}
Then, assuming that the solution of (\ref{ap3}) has the form
\begin{equation}
\displaystyle {\psi(x,t) = f(x)
\exp\Big(i \, \omega \, t \Big)}
\end{equation}
we find the eigenvalue
problem for
$f(x)$,
\begin{equation}
-\frac{\partial^{2} f}{\partial x^{2}} +
\Big[1-\frac{2}{\cosh^{2}(x)}\Big] f =
\omega^{2} f.
\label{ap4}
\end{equation}
This equation admits the following eigenfunctions with their respective
eigenvalues
\begin{eqnarray}
f_{T}(x) = \frac{2}{\cosh(x)}, \,\ \omega_{T}^{2} & = & 0, \\
f_{k}(x) = \frac{\exp(i k x) \,\ [k + i \mbox{tanh}(x)]}{\sqrt{2 \pi}
\,\ \omega_{k}}, \,\
\omega_{k}^{2} & = & 1 + k^{2},
\label{ap5}
\end{eqnarray}
which represent, respectively, the translation (Goldstone) mode and 
the radiation modes. 
Importantly,
the functions $f_{T}(x)$ and $f_{k}(x)$ form a complete set with the
orthogonality relations
\begin{eqnarray}
\int_{-\infty}^{+\infty} f_{T}^{2}(x) \,\ dx & = & 8, \quad\\
\int_{-\infty}^{+\infty} f_{T}(x) f_{k}(x) \,\ dx &=& 0, \quad \\
\int_{-\infty}^{+\infty} f_{k}(x) f_{k'}^{*}(x) \,\ dx &=& \delta(k-k').
\label{ap6}
\end{eqnarray}

We can now proceed with our problem:
In order to tackle Eq.\ (\ref{ecua1}), with noise as given 
in (\ref{ecua2}), 
we use the same {\em Ansatz}
proposed for the overdamped case in \cite{nos} 
(or for the general Klein-Gordon 
system in \cite{raja}): We assume that the solution of Eq.\ (\ref{ecua1}) is 
\begin{equation}
\phi(x,t)=\phi_{0}[x-X(t)] + \int_{-\infty}^{+\infty} dk A_{k}(t) f_{k}[x-X(t)],
\label{ecua3}
\end{equation}
where $X(t)$ is the kink position. 
We now insert (\ref{ecua3}) in (\ref{ecua1}) and
use the orthogonality of $f_{k}$ and $f_{T}$ \cite{bis}, obtaining
the following system of differential equation for $X(t)$ and $A_{k}(t)$:

\onecolumn 

\begin{eqnarray}
\ddot{X}(t)+\alpha \dot{X}(t) & = &   
- \frac{\alpha}{8} \dot{X}(t) \int_{-\infty}^{+\infty} dk A_{k}(t) I_{1}(k) - 
\frac{1}{16} \int_{-\infty}^{+\infty} dk \int_{-\infty}^{+\infty} dk' A_{k}(t) A_{k'}(t) R_{3}(k,k') +  \nonumber \\
& + & \frac{\sqrt{D}}{8} \int_{-\infty}^{+\infty} f_{T} [x - X(t)] \,\ \eta(x,t) \,\ dx - \nonumber \\
& - & \frac{1}{48}
\int_{-\infty}^{+\infty} dk \int_{-\infty}^{+\infty} dk_1 \int_{-\infty}^{+\infty} dk_2 A_{k}(t) A_{k_{1}}(t) A_{k_{2}}(t) 
R_{6}(k,k_{1},k_{2}) - \nonumber \\ 
& - & \frac{\dot{X}(t)}{4}\int_{-\infty}^{+\infty} dk 
\frac{\partial{A_{k}}}{\partial{t}}I_{1}(k) - 
\frac{\ddot{X}(t)}{8}\int_{-\infty}^{+\infty} dk A_{k}(t) I_{1}(k) + \nonumber \\
& + & \frac{\dot{X}^{2}(t)}{8}\int_{-\infty}^{+\infty} dk 
\frac{\partial{A_{k}}}{\partial{t}}I_{2}(k)
,\label{ecua4}
\end{eqnarray}

\begin{eqnarray}
\frac{\partial^{2}{A_{k}}}{\partial{t}^{2}}+
\alpha \frac{\partial{A_{k}}}{\partial{t}} + \omega_{k}^{2} A_{k}(t) & = &   
\alpha \dot{X}(t) \int_{-\infty}^{+\infty} dk' A_{k'}(t) I_{3}(k',k) + \nonumber \\
& + & \frac{1}{2} \int_{-\infty}^{+\infty} dk \int_{-\infty}^{+\infty} dk'
A_{k}(t) A_{k'}(t) R_{4}(k,k') - \nonumber \\ 
& - & \sqrt{D} \int_{-\infty}^{+\infty} 
f_{k}^{*} [x - X(t)] \,\ \eta(x,t) \,\ dx + \nonumber \\ 
& + & \frac{1}{6} 
\int_{-\infty}^{+\infty} dk' \int_{-\infty}^{+\infty} dk_1 \int_{-\infty}^{+\infty} dk_2 A_{k'}(t) A_{k_{1}}(t) 
A_{k_{2}}(t) R_{7}(k',k,k_{1},k_{2})+\nonumber \\ 
& + & 2 \dot{X}(t) \int_{-\infty}^{+\infty} dk' \frac{\partial{A_{k'}}}{\partial{t}}I_{3}(k',k) +  \nonumber \\ 
& + & \ddot{X}(t) \int_{-\infty}^{+\infty} dk' A_{k'}(t) I_{3}(k',k) + 
\dot{X}^{2}(t) I_{1}(k), 
\label{ecua5}
\end{eqnarray}
where 
\begin{eqnarray}
I_{1}(k) & = & \int_{-\infty}^{+\infty} 
\frac{\partial f_{k}}{\partial \theta} f_{T}(\theta) d\theta = 
\frac{i \pi \omega_{k}}{\sqrt{2 \pi} \,\ \cosh\Big(\displaystyle{\frac{\pi k}{2}}\Big)}, \nonumber \\
I_{2}(k) & = & \int_{-\infty}^{+\infty} 
\frac{\partial^{2} f_{k}}{\partial \theta^{2}} f_{T}(\theta) d\theta, \nonumber \\
R_{3}(k,k') & = & \int_{-\infty}^{+\infty} f_{T}(\theta) 
\frac{\partial f_{T}}{\partial \theta} f_{k}(\theta) 
f_{k'}^{*}(\theta) d\theta =  
-\frac{i (\omega_{k}^{2} - \omega_{k'}^{2})^{2}}{4 \omega_{k} \omega_{k'} 
\,\ \sinh \Big(\displaystyle{\frac{\pi \Delta k }{2}} \Big)},\,\ \Delta k=k'-k, \nonumber \\
I_{3}(k,k') & = & \int_{-\infty}^{+\infty} 
\frac{\partial f_{k}}{\partial \theta} f_{k'}^{*}(\theta) d\theta, \nonumber \\
R_{4}(k,k') & = & \int_{-\infty}^{+\infty} [f_{k'}^{*}(\theta)]^{2}   
\frac{\partial f_{T}}{\partial \theta} f_{k}(\theta) 
 d\theta, \,\ R_{4}(k,k) = \frac{3 i \omega_{k}}{8 \sqrt{2 \pi} \,\ 
\cosh \Big(\displaystyle{\frac{\pi k}{2}} \Big)}, \nonumber \\
R_{6}(k,k_{1},k_{2}) & = & \int_{-\infty}^{+\infty} 
\frac{\partial^{2} f_{T}}{\partial \theta^{2}} f_{k}(\theta) f_{k_1}^{*}(\theta) 
f_{k_2}(\theta) d\theta, \,\ \nonumber \\
R_{7}(k,k',k_{1},k_{2}) & = & \int_{-\infty}^{+\infty} \cos(\phi_{0}) 
f_{k'}^{*}(\theta) f_{k}(\theta) f_{k_1}^{*}(\theta) f_{k_2}(\theta)
 d\theta.
\label{ecua6}
\end{eqnarray}

It goes without saying that these equations can not be solved. Therefore, 
in order to extract information from them, we resort to a perturbative 
approach assuming the noise term is small, or equivalently, that the 
temperature and the dissipation
are not too large (this is not a serious restriction since 
our single-kink approach does not apply to high temperatures, when 
kink-antikink pairs are thermally generated \cite{butt}). We then 
expand $X(t)$ and $A_{k}(t)$  
in powers of $\sqrt{D}$, i.e., 
$X(t) = \sum_{n=1}^{\infty} (\sqrt{D})^{n} X_{n}(t)$ and 
$A_{k}(t) = \sum_{n=1}^{\infty} (\sqrt{D})^{n} A_{k}^{n}(t)$, since when 
$\sqrt{D}=0$ and $\alpha=0$ we recover the static kink solution 
(in this case initially centered at the origin) of the sG equation.   
By substituting these expansions in (\ref{ecua4}) and (\ref{ecua5}) 
we find a set of linear 
equations for the coefficients of these series. The first members of
this hierarchy correspond to 
order {$O(\sqrt{D})$}:
\begin{equation}
\ddot{X}_{1}(t) + \alpha \dot{X}_{1}(t)  =  \frac{1}{8}
\int_{-\infty}^{+\infty} f_{T} [x - X(t)] \,\ \eta(x,t) \,\ dx \equiv
\epsilon_{1}(t), 
\label{pez}
\end{equation}
from where we obtain the statistical properties of $\epsilon_{1}(t)$,
\begin{eqnarray}
 \langle\epsilon_{1}(t) \rangle=0, \quad
 \langle\epsilon_{1}(t) \epsilon_{1}(t')\rangle = \frac{1}{8} \delta(t-t'), 
 \label{ecua7}
 \end{eqnarray}
and 
\begin{equation}
\frac{\partial^{2} {A_{k}^{1}}}{\partial t^{2}}(t) + 
\alpha \frac{\partial {A_{k}^{1}}}{\partial t}(t) + 
\omega_{k}^{2} A_{k}^{1}(t) = \int_{-\infty}^{+\infty}
f_{k}^{*} [x - X(t)] \,\ \eta(x,t) \,\ dx\equiv\xi_{k}(t), 
\end{equation}
which in turn leads to 
 \begin{eqnarray}
\langle\xi_{k}(t) \rangle=0, \quad
\langle\xi_{k}(t) \xi_{k'}(t')\rangle = \delta(t-t') 
\delta(k-k'). 
\label{ecua8}
\end{eqnarray} 

Equations (\ref{pez})--(\ref{ecua8}) have been obtained in \cite{sal} by 
using a similar, but more restrictive perturbative approach. 
By integrating these two equations we obtain the 
first-order terms, $X_{1}(t)$ and $A_{k}^{1}$:
\bq
\ba{l}
\dst X_{1}(t) =  \int_{0}^{t} e^{-\alpha t'} 
\int_{0}^{t'} e^{\alpha \tau} \epsilon_{1}(\tau) d \tau \, dt', \quad 
\dst A_{k}^{1}(t) = e^{-\xmed{\alpha t}}
\left\{ C_1(t) \sin \omega t + C_2(t) \cos \omega t
\right\}, \\
\\
\dst 
C_1(t)= \frac{1}{\omega} 
\int_{0}^{t}   \xi_{k}(\tau) e^{\xmed{\alpha\tau}} \cos 
\omega\tau  d \tau, \quad
C_2(t)=- \frac{1}{\omega} 
\int_{0}^{t}   \xi_{k}(\tau) e^{\xmed{\alpha\tau}} \sin 
\omega\tau  d \tau, 
\label{ecua14} 
\ea
\eq
where $\omega^2 =\omega_k^2 -(\alpha^2/4)$. 
{}From these relations 
we can calculate the mean values and correlation functions up to 
first order in $\sqrt{D}$:
\begin{eqnarray}
\langle X_{1}(t) \rangle  =  0, \quad \dst \langle X(t) X(t')\rangle & = &  
D \langle X_{1}(t) X_{1}(t')\rangle 
 = \frac{D}{16 \alpha^{3}} \Big[
 e^{-\alpha M}-e^{-\alpha|\Delta t|} + e^{-\alpha M -\alpha|\Delta t|} - \nonumber \\
& - & e^{-\alpha(t+t')} +  e^{-\alpha t}+e^{-\alpha t'} +
 2(\alpha M -1)\Big],
 \label{ecua15}
 \end{eqnarray}
\begin{eqnarray} 
\dst 
\langle\dot{X}_{1}(t) \rangle  & = & 0, \quad\quad  
\langle\dot{X}(t) \dot{X}(t')\rangle = D
 \langle\dot{X}_{1}(t) \dot{X}_{1}(t')\rangle = 
\frac{D}{16 \alpha}  \Big[ 
 e^{-\alpha |\Delta t|} - e^{-\alpha(t+t')} \Big], 
\label{ecua16}
 \end{eqnarray}
\begin{eqnarray}
\langle A_{k}^{1}(t) \rangle = 0, \quad \langle A_{k}(t) A_{k}(t')\rangle & = &  
D \langle A_{k}^{1}(t) A_{k}^{1}(t')\rangle = 
\frac{D}{ \omega^{2}}  e^{-\alpha(t+t')/2}
\Big[ \frac{e^{\alpha M}-1}{2\alpha} \cos \omega\Delta t - \nonumber \\
& - & \frac{\alpha e^{\alpha M}}{8\omega_k^2 } \cos \omega\Delta t - 
\frac{\omega e^{\alpha M}}{4\omega_k^2 } \sin \omega |\Delta t| 
+ \frac{ \alpha }{8\omega_k^2 } \cos\omega(t+t')- \nonumber \\
& - &  \frac{ \omega }{4\omega_k^2 } \sin\omega(t+t')\Big], 
\label{ecua17}
\end{eqnarray}
where $\Delta t=t-t'$, and $M=\min(t,t')$. Of course, 
for $t'=t$ in Eq.\ (\ref{ecua15}) we recover the result in \cite{kaup} for 
$\langle [X(t)]^2 \rangle $.

We now turn to the main point of our work, 
namely obtaining 
the next-order corrections for the position 
and the velocity of the center of the kink. This requires the calculation
of the next two contributions to $X(t)$ as well as the second order 
in the radiation terms, which are:\bigskip

\underline{$O(D)$}
\begin{eqnarray}
\displaystyle \ddot{X}_{2}(t) + \alpha \dot{X}_{2}(t) & = & \epsilon_{2}(t), 
\label{ecua9}
\end{eqnarray}
\begin{eqnarray}
\displaystyle 
\epsilon_{2}(t) & \equiv & -\frac{\epsilon_{1}(t)}{8} \int_{-\infty}^{+\infty} dk A_{k}^{1}(t) I_{1}(k) - 
\frac{\dot{X}_{1}(t)}{4} \int_{-\infty}^{+\infty} dk \frac{\partial {A_{k}^{1}}}{\partial t}(t) I_{1}(k) - \nonumber \\
& - & \frac{1}{16} \int_{-\infty}^{+\infty} dk' 
\int_{-\infty}^{+\infty} dk 
A_{k}^{1}(t) A_{k'}^{1}(t) R_{3}(k,k'), 
\label{ecua10}
\end{eqnarray}

\begin{eqnarray}
\displaystyle 
\frac{\partial^{2} {A_{k}^{2}}}{\partial t^{2}}(t) + 
\alpha \frac{\partial {A_{k}^{2}}}{\partial t}(t) + \omega_{k}^{2}  
A_{k}^{2}(t) & = & \epsilon_{1}(t) 
\int_{-\infty}^{+\infty} dk' A_{k'}^{1}(t) I_{3}(k',k)  + \nonumber \\
& + & \frac{1}{2} 
\int_{-\infty}^{+\infty} dk \int_{-\infty}^{+\infty} dk' A_{k}^{1}(t) A_{k'}^{1}(t) R_{4}(k,k') - \nonumber \\ 
& - & \dot{X}_{1}^{2}(t)  I_{1}(k) + 2 \dot{X}_{1}(t) \int_{-\infty}^{+\infty} dk'  
\frac{\partial {A_{k'}^{1}}}{\partial t}(t) I_{3}(k',k); 
\label{ecua11}
\end{eqnarray}

\underline{$O([\sqrt{D}]^{3})$}
\begin{eqnarray}
\displaystyle
\ddot{X}_{3}(t) + \alpha \dot{X}_{3}(t) & = & \epsilon_3(t), 
\label{ecua12} 
\end{eqnarray} 
\begin{eqnarray}
\displaystyle
\epsilon_3(t) & \equiv &
- \frac{\epsilon_{1}(t)}{8} \int_{-\infty}^{+\infty} dk A_{k}^{2}(t) I_{1}(k) - 
\frac{\epsilon_{2}(t)}{8} \int_{-\infty}^{+\infty} dk A_{k}^{1}(t) I_{1}(k) - \nonumber \\
& - & \frac{1}{16} \int_{-\infty}^{+\infty} dk \int_{-\infty}^{+\infty} dk' A_{k}^{2}(t) A_{k'}^{1}(t) R_{3}(k,k') - \nonumber \\
& - & \frac{1}{16} \int_{-\infty}^{+\infty} dk \int_{-\infty}^{+\infty} dk' A_{k}^{1}(t) A_{k'}^{2}(t) R_{3}(k,k') - \nonumber \\
& - & \frac{1}{48} \int_{-\infty}^{+\infty} dk \int_{-\infty}^{+\infty} dk_{1} 
\int_{-\infty}^{+\infty} dk_{2} A_{k}^{1}(t) 
A_{k_{1}}^{1}(t) A_{k_{2}}^{1}(t) R_{6}(k,k_{1},k_{2})+ \nonumber \\
& + & \frac{\dot{X}_{1}^{2}(t)}{8} \int_{-\infty}^{+\infty} dk \frac{\partial {A_{k}^{1}}}{\partial t}(t) I_{2}(k)- 
\frac{\dot{X}_{1}(t)}{4} \int_{-\infty}^{+\infty} dk \frac{\partial {A_{k}^{2}}}{\partial t}(t) I_{1}(k)-\nonumber \\
& - & \frac{\dot{X}_{2}(t)}{4} \int_{-\infty}^{+\infty} dk \frac{\partial {A_{k}^{1}}}{\partial t}(t) I_{1}(k).
\label{ecua13} 
\end{eqnarray} 

Analogously to what we have done for 
the first-order corrections, from the solutions of 
Eqs.\ (\ref{ecua9}) and (\ref{ecua12}) we find that   
\begin{eqnarray}
\langle X_{2}(t) \rangle & = & 0, \,\ \langle\dot{X}_{2}(t) \rangle = 0, \label{ecua18}\\
\langle X_{3}(t) \rangle & = & 0, \,\ \langle\dot{X}_{3}(t) \rangle = 0. 
\label{ecua19}
\end{eqnarray}
As for higher moments, taking into account that the cross-correlation 
function of $X_{1}(t)$ and 
$X_{3}(t')$ is of the same order as $\langle X_{2}(t) X_{2}(t') \rangle$, 
and also 
that $\langle X_{1}(t) X_{2}(t') \rangle=0$ we obtain that  

\begin{eqnarray}
\langle[X(t)]^{2}\rangle & = & 
D \langle[X_{1}(t)]^{2}\rangle + \nonumber \\
& + & 
D^{2} (\langle[X_{2}(t)]^{2}\rangle + 2 
\langle X_{1}(t) X_{3}(t) \rangle )+...,
\label{ecua20}
\end{eqnarray}
\begin{eqnarray}
\langle[\dot{X}(t)]^{2}\rangle & = & 
D \langle[\dot{X}_{1}(t)]^{2}\rangle + \nonumber \\
& + & 
D^{2} (\langle[\dot{X}_{2}(t)]^{2}\rangle + 2 
\langle\dot{X}_{1}(t) \dot{X}_{3}(t) \rangle )+...   
\label{ecua21}
\end{eqnarray}

The expressions for the functions $\langle[X_{2}(t)]^{2}\rangle$,  
$\langle[\dot{X}_{2}(t)]^{2}\rangle$, $\langle X_{1}(t) X_{3}(t) \rangle$,
and $\langle\dot{X}_{1}(t) \dot{X}_{3}(t) \rangle$ 
can be obtained after a lengthy calculation, and 
are very cumbersome indeed. We therefore do not
include them here. However, for large time ($t >> 1/\alpha$) 
these relations can be simplified, yielding, as $t\to\infty$ 

\twocolumn

\begin{eqnarray}
\langle[X(t)]^{2}\rangle & = & \frac{ k_{b} T t}{4 \alpha} \Big\{1 + 
\frac{k_{b} T}{32} \Big(1 + 
\frac{9 \sigma^{2}}{4} \Big) \Big\}, 
\label{ecua22}
\end{eqnarray}

\begin{eqnarray}
\langle[\dot{X}(t)]^{2}\rangle & = & \frac{ k_{b} T}{8} \Big\{1 +  
\frac{3 k_{b} T}{128} \Big(12 + \sigma^{2} \Big) \Big\}, 
\label{ecua23}
\end{eqnarray}
with
\begin{equation}
{ \sigma = \int_{-\infty}^{+\infty} \frac{dk}{\omega_{k} 
\cosh\Big(\displaystyle{\frac{\pi k}{2}}\Big)}} = 1.62386. 
\end{equation}

To complete the characterization of the kink diffusion, we can now
compute in a straightforward way the
average value of the wave function $\phi(x,t)$, defined as

\begin{eqnarray}
\nonumber
\langle \phi(x,t)\rangle &=&\langle \phi_{0}[x- \sqrt{D} X_1(t)]\rangle + 
O(D) =\\ &=&  \int_{-\infty}^{+\infty} 
dX_{1} p(X_{1}) \phi_{0}[x-\sqrt{D} X_{1}(t)],
\label{ecua24} 
\end{eqnarray}
where $p(X_{1})$ is the probability distribution function for $X_{1}$. 
To find explicitly this function we note that, if we rewrite 
Eq.\ (\ref{pez}) as a system of two differential equations,

\begin{eqnarray}
\dot{X}_{1} & = & V, \nonumber \\
\dot{V} & = & - \alpha V + \epsilon_{1}(t),
\label{ecua25} 
\end{eqnarray}
the last equation represents an Ornstein-Uhlenbeck process for the velocity,
and its distribution function is given by 

\begin{equation}
p(V)=\sqrt{\frac{1}{2 \pi\langle V^2\rangle}} 
\exp\Big(- \frac{V^{2}}{2\langle V^2\rangle}  \Big), 
\label{pdv}
\end{equation}
(see \cite{noise}). Subsequently, 
by integrating the first equation of (\ref{ecua25}), 
we obtain that $X_{1} = \int_{0}^t V(\tau) d \tau$. Since $V$ has
a Gaussian distribution function, $X_{1}$ has also a Gaussian 
distribution function, given by (recall that $\langle X_1(t)\rangle=0$)

\begin{equation}
{\displaystyle{ 
p(X_{1})=\sqrt{\frac{1}{2 \pi \langle[X(t)]^{2}\rangle }} 
\exp\Big(-\frac{1}{2} \frac{X_{1}^{2}}{\langle[X(t)]^{2}\rangle} 
\Big) }}, 
\label{pdx}
\end{equation}
where the first and second moments of $X_{1}$ 
were obtained before, see Eq.\ (\ref{ecua15}).
With this result, 
the integral (\ref{ecua24}), can be evaluated numerically 
taking into account Eqs.\ (\ref{pdx}) and  
(\ref{ecua15}). In the next section we will compare this result with the 
mean value of the wave function as obtained from 
simulations of the full partial differential equation (\ref{ecua1}). 

\section{Numerical simulations}
 
In order to test the approximate theory developed in the previous section,
we have simulated numerically Eq.\ (\ref{ecua1}) by using 
the Heun method \cite{maxra}. In our simulations we begin with a kink, 
initially 
at rest, with free boundary conditions. For the damping 
coefficient we choose $\alpha=0.1$, which is not too small because from 
(\ref{ecua22}) we can see that $\langle[X(t)]^{2}\rangle$ 
is proportional to $1/\alpha$. This means that if $\alpha$ is too 
small the kink can move in a much larger region, forcing us 
to increase the length of our simulated system in the simulations,
already quite time consuming. 
Furthermore, the relation (\ref{ecua22}) is only valid for large  
times ($t >> 1/\alpha$). Again, for too small $\alpha$ we would 
need to simulate 
our equation for very long times and,
as $\langle[X(t)]^{2}\rangle$ increases
linearly with time [see Eq.\ (\ref{ecua22})], 
the system length would once more have to be large. 
The other parameters are
$\Delta x = 0.2$, $\Delta t = 0.001$  and the length of the system $L=200$. 
We have calculated all average values over 1000 realizations 
up to a final time $400$. It is important to point out that, this system
being inertial, the accuracy of the averages is considerably less than 
for overdamped problems, this being the reason why we have to use such 
large ensembles of trajectories to obtain reasonably good results. 

\begin{figure}
\begin{center}
\epsfig{file=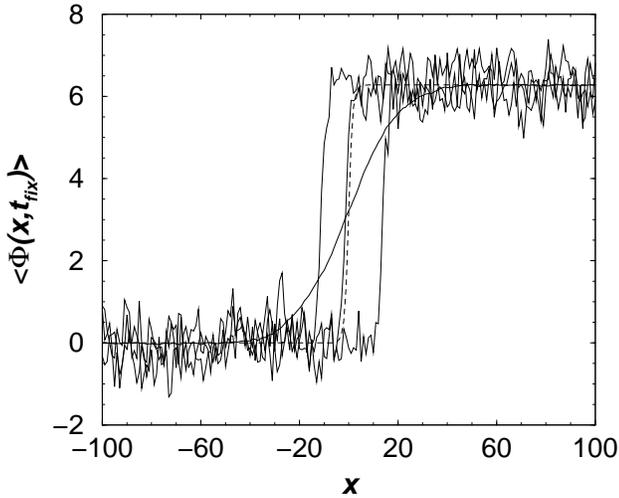, width=2.7in, angle=-90}
\caption[]{
{}Individual kink realizations compared with
initial conditions and averages. Shown 
are three individual realizations at $t=300$ with parameters $k_BT=0.4$ 
and $\alpha=0.1$ (thin solid lines), the initial condition given by a kink
at rest (dashed line) and the mean value of $\phi$ at the same time obtained
from averaging over 1000 realizations (thick solid line).
}  
\label{graph0}
\end{center}
\end{figure}

An important, nontrivial issue is the question 
as to how can we
find the center of the kink. We solve this problem by finding 
all the discrete lattice points
$x_{i}$ and  $x_{i+1}$ such that  
$\phi_{i} \le \pi$ and $\phi_{i+1} \ge \pi$ or vice versa, 
and then estimating the corresponding points
$\tilde{x}_{i}$ where $\phi=\pi$ by linear
interpolation. Afterwards, among the 
$n$ such points $\tilde{x}_{n}$, 
we choose to be the center of the kink the value $\tilde{x}=\tilde{x}_{n}$, 
which minimizes 
$\sum_{i=1}^{L/\Delta x}[\phi_{i}(t) - \phi_{0}(x-\tilde{x}_{n},t)]^2$,
i.e., the discrete version of the integral of the square of the difference
of $\phi$ and $\phi_0$. 
It has to be realized that
this involves an assumption, namely that individual realizations of the kink
have a shape similar to that of the unperturbed kink. As can be seen from 
Fig.\ \ref{graph0}, where the individual realizations are compared with 
the initial condition (an exact kink), this is indeed the case and our 
procedure is truly sensible. Therefore, we are sure that 
this method to compute the kink center 
avoids the false centers, which can appear for higher temperatures due to  
fluctuations introducing a systematic difference 
between the numerical and the theoretical 
results (see \cite{nos}). With the procedure we have just summarized, that
works even for relatively large temperatures, we
believe we find a very accurate approximation to the actual center of the
kink. We will come back to Fig.\ \ref{graph0} below. 

\begin{figure}
\begin{center}
\epsfig{file=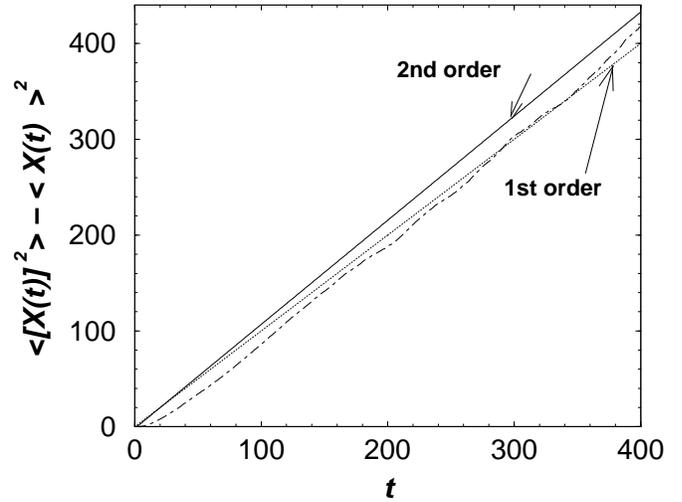, width=2.7in, angle=-90}
\caption[]{
{}Comparison of analytical and numerical results for the 
variance of the kink center, 
$\langle [X(t)]^{2}\rangle - \langle X(t)\rangle ^2$, 
for $k_{b} T=0.4$. The dot-dashed line is the result of the 
numerical simulation of Eq.\ (\ref{ecua1}), whereas the dotted and the 
solid lines are the analytical results for the first- (dotted line)
and second-order (solid line) expressions 
[Eqs.\ (\ref{ecua15}) and 
(\ref{ecua22}) with $t=t' $ respectively]. Only the slopes of the lines 
for $t >> 10$ have to be compared.    
}  
\label{graph1}
\end{center}
\end{figure}

\begin{figure}
\begin{center}
\epsfig{file=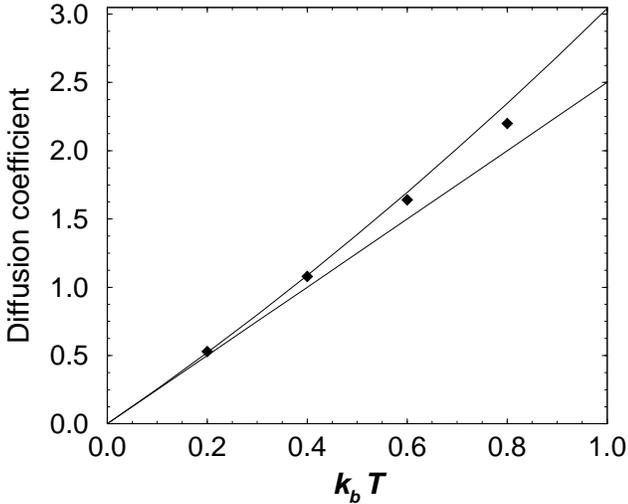, width=2.7in, angle=-90}
\caption{Kink diffusion constant vs temperature. 
The solid lines represent the analytical diffusion 
coefficient up to first- (lower line)
and second-order (upper line). 
Diamonds stand for numerical values of 
the kink center diffusion coefficient,
obtained by numerical integration of Eq.\ (\ref{ecua1}).
}
\label{graph2}
\end{center}
\end{figure}

As an example of the comparison of 
the numerical simulations of Eq.\ (\ref{ecua1}) with the theoretical results
obtained in the previous section and valid for large times, 
Fig.\ \ref{graph1}
shows the numerically computed variance of the center of the kink,
 $\langle[X(t)]^{2}\rangle - \langle X(t) \rangle^{2}$,
as well as the first- and second-order analytical expressions. 
The plot clearly evidences that the numerical variance 
asymptotically coincides
with the second-order expression: Note that to compare the different 
curves one has to look at the {\em slopes} at times 
$t >> 1/\alpha$ (in this case, 
$t \ge 100$, for instance, as $\alpha=0.1$);
the theoretical result is
not valid at early times and therefore there is a bias between 
analytics and numerics coming from that. The small, irregular 
oscillations in the numerical curve arise from the difficulty in 
accurately computing averages in an underdamped system like this 
mentioned above; however, we believe that the present accuracy is 
enough to confirm the validity of our approach. 
We have observed the same agreement 
for other values of temperatures ($k_{b} T = 0.2, 0.6, 0.8$, not shown). 
In all cases, 
we have computed the diffusion coefficient 
for large times as the slope of the variance of $X(t)$ 
again for 
$t \ge 100$, the regime in which we expect
our analytical approximation to be valid. 
Summarizing our results, these numerical values of the diffusion 
coefficient are plotted in Fig.\ \ref{graph2} together with the 
theoretical results. It is 
clear that for large temperatures the quadratic behavior in $k_{b} T$ of the 
diffusion coefficient becomes important. For higher values of the 
temperature, such as $k_{b} T = 0.8$,
the numerical 
value of the diffusion coefficient is not so close to the predicted one. 
This effect arises because of the large diffusivity of 
the kink in that range: Indeed, for this and higher temperatures the kink 
performs very long excursions away from the center, reaching the boundaries
of the numerical integration interval; it is clear that when this occurs, 
the diffusion of the kink is not in free space anymore and hence those 
realizations spoil the quality of the averages. The way to solve this 
problem would be to resort to much larger numerical systems, but within 
our present computing capabilities this would necessitate a simultaneous
decrease in the number of realizations in the average, leading again to
poorer results. However, it is important to realize that this boundary 
effect leads to an {\em underestimation} of the diffusion coefficient 
(as the boundary prevents the kink from travelling as far as it should)
and therefore the point in Fig.\ \ref{graph2} for $k_{b} T = 0.8$ is 
a lower bound for the diffusion coefficient, with the actual one lying
even closer to our second order prediction.

\begin{figure}
\begin{center}
\epsfig{file=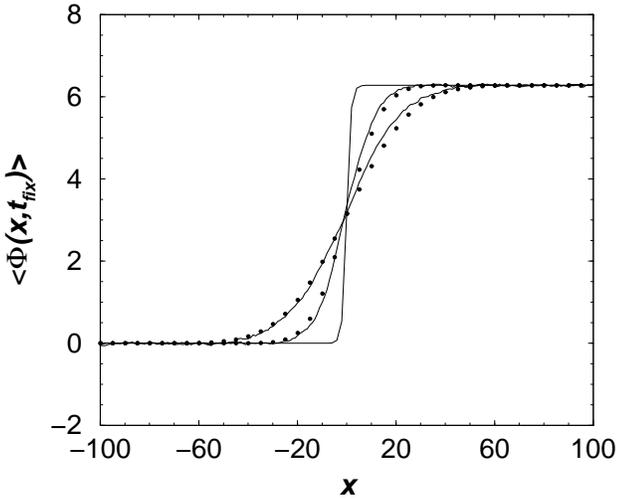, width=2.7in, angle=-90}
\caption{Mean value of the 
wave function (solid line) for two fixed times, $t_{fix}=100$ and 
$t_{fix}=300$, with $k_{b} T = 0.6$, obtained from 
numerical simulations of Eq.\ (\ref{ecua1}) 
compared to
the values of $\langle \phi(x,t)\rangle$ (points)
obtained numerically from the integral in Eq.\ (\ref{ecua24}). 
The narrowest solid line is the initial data (kink initially at rest). 
}
\label{graph3}
\end{center}
\end{figure}

Finally, there is one last question that deserves discussion, namely that
of the
physical significance of the mean value of the wave
function $\phi$. In Fig.\ \ref{graph0} we can clearly see that, whereas 
individual realizations of kinks look very similar to the unperturbed ones,
the mean value of $\phi$ is a much wider excitation, not even close to the
original kink. Figure \ref{graph0} clearly shows that
this does not mean that the width of individual kinks increases;
indeed, much as we discussed regarding the overdamped problem \cite{nos}, 
we have verified numerically that 
the mean wave function $\langle \phi(x,t_{fix}) \rangle$ increases due to the 
variance of the kink center of individual realizations, and hence it should
not be interpreted as the typical deformation of the shape of kinks:
Indeed, the widening of the mean value of $\phi$ arises from
the contributions of the stochastically moving, but mostly undistorted kinks
whose center positions have the distribution of a rigid, diffusing particle. 
To further check this interpretation, 
we can look at Fig.\ \ref{graph3},
where we have represented the mean value of the wave function 
for two fixed times $t_{fix}=100, 300$, obtained from the numerical simulation 
of the full partial differential equation (\ref{ecua1}),
for $k_{b} T = 0.6$ and $\alpha = 0.1$. The overimposed points,
computed by using the Gaussian distribution 
function $p(X_{1})$ [Eq.\ (\ref{pdx})] of the kink center $X(t) = \sqrt{D}
X_{1}$ found in the last section, show the excellent agreement between 
our theory and the simulation. Of course, there is a small discrepancy that
is likely to disappear if one would go to a next order calculation, but for
the present purpose of understanding the mean wave function $\phi$ the first
order calculation is enough. In addition, 
we have plotted the initial kink (at rest) in order to see that the 
mean value of the wave function increases with time. 

\section{Discussion and conclusions}

To summarize,
in this work we have studied the diffusive dynamics of sine-Gordon kinks 
subjected to thermal fluctuations. We have analytically calculated 
expressions, valid up to second order in temperature, for the average 
position and variance of the kink center, as well as for the mean shape
of the kink. We have numerically checked the validity of these results
up to temperatures of the order of $k_bT=0.8$ (in dimensionless units,
equivalent to about a 10\% of the kink rest mass), already close to the 
temperature at which kink-antikink nucleation becomes a likely event.
Therefore, our first conclusion is that the second-order theory 
developed here is the proper one, meaning it is accurate and higher 
order terms are negligible, to describe the thermal diffusion of 
sine-Gordon kinks in the single kink propagation regime. 
Interestingly, our calculation pinpoints the fact that the second-order
correction in $k_{b} T$ comes from the interaction between kink and phonons. 
This implies that the physics behind this contribution is basically the
same as for the case of anomalous 
diffusion in an isolated chain mentioned in the
introduction \cite{yuripr,Wsch,March2,March}. Note that we do not expect
$T^{-1}$ contributions in our analytical calculations, as they are 
carried out in a continuum sG equation \cite{yuripr} and, in any case,
they would show up in simulations only for very low temperatures. 
Apart from that,
it is also interesting to note that, according to Eq.\ (\ref{ecua23}), 
the second order term implies an increase of the energy carried by the
kink beyond the $k_bT/2$ predicted by statistical mechanics (recall that
the kink mass is 8 in our units). This can be interpreted in the following
way: The kink is dressed by phonons which increase its mass. Thus, the 
kink energy would be $M(T)\cdot\dot{X}^2/2$, with a temperature
dependent mass $M(T)$ whose expression can be easily found from 
Eq.\ (\ref{ecua23}). In order to confirm this interpretation, one could
compute the energy carried by the phonons which dress the kink, but we
believe that it is not necessary because, on the one hand, it would be
a rather involved calculation (far beyond the scope of this work) and, on
the other hand, we do not think that there is any other possible 
interpretation of this result. 

A second relevant point of this study relates to the numerical simulation
and center location procedures. As this is an underdamped (inertial) 
system, the thermal mobility of kinks is quite large, the larger the 
higher the temperature. Because of this, we have not been able to obtain
very precise numerical averages at the top of the temperature range 
studied, since the lengths of the systems and the number of realizations
required are very large and consequently time consuming. However, we 
believe that the results presented here are enough to verify our theory.
This is reinforced by the very good agreement between analytics and 
numerics regarding the mean shape of the field, even for temperatures
as large as $k_bT=0.6$ (see Fig.\ \ref{graph2}), which shows that our approach 
indeed captures the physics of the diffusion process. In addition,
we want to emphasize that, to our knowledge, we have designed a new 
algorithm to detect the kink center which gives very good results even
for the highest temperatures studied, where previous researches, such
as \cite{nos}, had found problems arising from the many false centers 
detected.

Another important issue is the comparison of the pre\-sent analysis to 
that in \cite{nos} for the overdamped problem. We have found that
the diffusion coefficient 
given by (\ref{ecua22}) for the present case practically coincides
with that obtained in \cite{nos} 
for the overdamped limit of the equation:
the difference in the second order is approximately $0.06 k_{b} T$, 
i.e., very small compared to the magnitude of the quadratic contribution
itself. Furthermore,
the width of the mean value of the wave function 
increases with time for the overdamped case \cite{nos}
in the same manner as that reported here.
Therefore, we can conclude 
that for large times the dynamic of the underdamped sG kinks is very
similar to the overdamped case. This is an important point, because in 
principle one can expect similar results for other kink-bearing systems
such as the $\phi^4$ equation, for instance, whose overdamped diffusive
dynamics is known (see \cite{dzi} for the $\phi^4$ case), thus avoiding
the much more involved calculation of the underdamped case. 

Finally, we want to mention the relevance of this work to experimental
systems, such as long Josephson junctions. As has been shown in 
\cite{exp}, the thermal sG equation (\ref{ecua1}) is 
a good description of the physics of in-line Josephson junctions 
(although different boundary conditions are needed in that case). 
The work in \cite{exp} compared the predictions from the sG model to
experimentally measured escape rates from the zero voltage state. 
Therefore, it should be possible to design similar experiments in 
order to test our results and, specifically, the increased (quadratic)
diffusivity of kinks at higher temperatures vs the linear behavior 
at lower ones. We hope that our theoretical work stimulates further
experimental research in that direction.

\section*{Acknowledgement}
We are grateful to A. R. Bishop for his comments and Esteban Moro for his 
suggestions.  
Work at GISC (Legan\'es) has been supported by CICyT (Spain) grant MAT95-0325
and DGES (Spain) grant PB96-0119. Travel between Bay\-reuth and Madrid is 
supported by ``Acciones Integradas Hispano-Alemanas'', a joint program of
DAAD (Az.\ 314-AI) and DGES. This research is part of a project supported 
by NATO grant CRG 971090.

\end{document}